# Understanding thermal induced escape mechanism of optically levitated sphere in vacuum

MENGZHU HU,[1] NAN LI,[1,4] ZHENHAI FU,[2] YIZHOU ZHANG,[2] WENQIANG LI,[1] HAN CAI,[1] AND HUIZHU HU[2,3,5]

[1]*College of Optical Science and Engineering, Zhejiang University, Hangzhou 310027, China*

[2]*Quantum Sensing Center, Zhejiang Lab, Hangzhou 310000, China*

[3]*State Key Laboratory of Modern Optical Instrumentation, Zhejiang University, Hangzhou 310027, China*

[4] *nanli@zju.edu.cn*

[5] *huhuizhu2000@zju.edu.cn*

**Abstract:** The escape phenomenon, mainly caused by thermal effects, is known as an obstacle to the further practical application of optical levitation system in vacuum. Irregular photophoresis induced by thermal effects can act as an "amplifier" of Brownian motion. Studies on this topic provide interpretation for particle escaping phenomenon during the pressure decreasing process, as well as valuable insights into the micro- and nanoscale thermal effects in optical trap in vacuum. In this paper, we derive and test a dynamic model for the motion of an optically levitated particle in a non-equilibrium state and demonstrate the escaping mechanism of heated particles. The result of theoretical investigations is consistent with experimental escape at 0.1mbar. This work reveals and provides a theoretical basis for the stable operation of laser levitated oscillator in high vacuum and paves the way for the practicability of ultra-sensitive sensing devices.

## 1．Introduction

Levitated particles in vacuum can be applied in a wide range of fields, including precision measurement of acceleration [1,2] and mass [3], ultrasensitive force [4,5] and torque detection [6,7], high-speed rotation [8,9], optical refrigeration [10], quantum ground-state cooling [11-15], and stochastic thermodynamics [16-18]. Unlike in liquid or air, optical tweezers operating in vacuum are well isolated from the thermal environment, making them an excellent candidate for ultrasensitive sensing. Interactions with the thermal environment cause the dissipation of the center-of-mass motion and are the source of random forces acting on the particles. However, the effects of laser heating are stronger in vacuum, since the heat exchange between particles and surroundings becomes insufficient with decreased pressure.

The thermal effects of a levitated particle have been suspected to be the cause of particle loss at decreased gas pressure in numerous researches [19-24]. Photophoretic force arising from the internal temperature gradient has proved to be the mechanism for the loss at ~30 mbar [24]. In this case, the particle is assumed to have a constant accommodation coefficient $\alpha$, and the photophoretic force is induced by variation in the temperature $T_s$ of the particle surface ($\Delta T_s$-force). The $\Delta T_s$-force is called space-fixed force because its direction is determined by the direction of the radiation and is almost independent of the orientation of the particle. As a matter of fact, there always exists a variation in accommodation coefficient over the surface of particle due to the impurities and non-ideal particle shape. This results in a $\Delta\alpha$-force on the particle from the location of higher accommodation to the location of lower one. The direction of this photophoretic force is determined by the orientation of the particle and is independent of the direction of the illumination[25], thus Δα-force is body-fixed. The direction of Δα-force varies with the orientation of the particle, which yields particle motion in any possible direction. As a result, random walk depending on the irradiation occurs, which adds to the Brownian motion [26].

In this paper, the motion of a heated trapped sphere is investigated. First, we study the dynamics of the sphere under both types of photophoretic forces in connection with Brownian motion. It is shown that irregular photophoresis due to the Δα-force enhances the stochastic process of Brownian motion. Then, we test the model by comparing the calculated results with the experimental data and previous work. The dynamic model allows us to assess the difference in accommodation coefficient over the levitated sphere, implying the application of levitodynamics to material science study. Since maintaining the trapping stability of levitated sphere is a critical task for an optically levitated system, our study paves the way for the stable operation of optomechanical oscillator in high vacuum.

## 2．Principle of photophoretic force

Photophoresis is a well-known phenomenon of the light-induced motion of particles suspended in gas [32]. There are two types of photophoretic force, namely, $\Delta T_s$-force $F_{\Delta T}$ and $\Delta \alpha$-force $F_{\Delta \alpha}$. These are respectively induced by variation in temperature over the surface of particles and by variation in the thermal accommodation coefficient $\alpha$. Both can cause a temperature variation in the gas surrounding the particle. After inelastic collision, hotter gas molecules leave the particle surface faster than colder ones, which results in a net force on the particle pointing from the hot to the cold side.

The $\Delta T_s$-force is directed along or against the direction of incident light (Fig.1(a)). A semi-empirical expression for $F_{\Delta T}$ has been given by Rohatschek [27] on spherical particles for the entire range of pressure $p$, such as:

$$F_{\Delta T} = F_{\max} \frac{2}{\frac{p}{p_{\max}} + \frac{p_{\max}}{p}}, \qquad (1)$$

$$p_{\max} = D\sqrt{\frac{2}{\alpha}} \frac{3T}{\pi a}, \quad F_{\max} = D\sqrt{\frac{\alpha}{2}} \frac{a^2 J_1}{k_p} I$$

$$D = \frac{\pi}{2}\sqrt{\frac{\pi}{3}\kappa} \frac{\bar{c}\eta}{T}, \quad \bar{c} = \sqrt{\frac{8}{\pi}\frac{RT}{M}}$$

where $\bar{c}$ denotes the average thermal velocity of gas molecules at temperature $T$, $M$ denotes the molar mass of the gas, $\eta$ represents the dynamic viscosity of gas, and $R$=8.31J/(mol·K) is the gas constant. The thermal creep coefficient $\kappa$ is related to the thermal accommodation coefficient $\alpha$; therefore, $D$ is a factor determined entirely by the gas properties, independent of the pressure $p$ and particle's radius $a$. $J_1$ represents the asymmetry parameter that involves an integration of normalized absorbed light intensity over the particle volume [28]. For a weakly absorbing sphere with a complex refractive index of $m = n + ik$ illuminated by a homogeneous plane wave at the light wavelength $\lambda$, $J_1$ can be obtained by the formula [29,30]:

$$J_1 = 2nkx\left(\frac{3(n-1)}{8n^2} - \frac{2}{5}nkx\right) \qquad (2)$$

where $x = 2\pi a/\lambda$ and $kx \ll 1$. It is also applicable to our studied configuration with the particle illuminated by focused laser beams with non-uniform intensity profiles, the intensity of the light field near the equilibrium position can be considered to be uniform. We assume that the light is +z-propagating, $J_1 < 0$ leads to positive photophoresis in the +z-direction, namely, the particle moves in the direction away from the radiation source. $I$ represents the flux density of illumination at the particle position, and $k_p$ is the thermal conductivity of the particle. The expression of $F_{\Delta T}$ can achieve a maximum force $F_{max}$ at a pressure $p_{max}$, where the particle size is comparable to the mean free path of gas molecules.

The particle can experience pure $\Delta T_s$-force only if the accommodation coefficient $\alpha$ is uniform. However, the accommodation coefficient $\alpha$ shall have variation over the particle surface, e.g., that arising from the difference in surface shape and roughness or different material composition of the particle. Therefore, even if the particle is heated evenly, there is still $\Delta\alpha$-force acting on the particle (Fig.1(a)). For a simple model in which the surface of spherical particle is divided into two hemi-spheres with two different accommodation coefficients $\alpha_1$ and $\alpha_2$, the expression for instantaneous $\Delta\alpha$-force is given by the following equation [28]:

$$F_{\Delta\alpha} = \frac{1}{2\bar{c}} \cdot \frac{\gamma-1}{\gamma+1} \cdot \frac{1}{1+(p/p^*)^2} \frac{\Delta\alpha}{\bar{\alpha}} \cdot H \tag{3}$$

where $\gamma = c_p/c_v$ represents the ratio of the specific heats of the gas, and $H$ denotes the net energy flux transferred by gas molecules. For a sphere in air, $\gamma = 1.4$ [31,32]:

$$F_{\Delta\alpha} = \frac{1}{12\bar{c}} \cdot \frac{1}{1+(p/p^*)^2} \frac{\Delta\alpha}{\bar{\alpha}} \cdot H \tag{4}$$

Here, $\Delta\alpha = \alpha_2 - \alpha_1$, $\bar{\alpha} = (\alpha_1+\alpha_2)/2$, and $p^* = \sqrt{\alpha/2}\, p_{max} = 3DT/\pi a$ is a characteristic pressure inversely proportional to the radius. The energy flux absorbed by the sphere is $H = Q_a \pi a^2 I$, and $Q_a$ is the absorption efficiency of the sphere [32]. The direction of $\Delta\alpha$-force ($F_{\Delta\alpha}$) points from the side of the higher accommodation coefficient to the lower one. This force is independent of the direction of incident light and is determined by the particle orientation, which is also called body-fixed force here. Since the effect of collisions between a particle and surrounding gas molecules can result in random force and torque, all particles perform Brownian displacement and Brownian rotation. As a result, the direction of $\Delta\alpha$-force is randomly distributed, which will make Brownian motion more vigorous. Illuminated particles may perform irregular photophoresis and exhibit an irregular motion shown in the insert of Fig.1(b), which is similar to Brownian motion but stronger. It is also obvious fom the formula (1) and (4) that photophoretic forces strongly depend on the pressure. The $\Delta T_s$-force reach its maximum at a pressure which enables the particle size comparable to the mean free path of gas molecules. The $\Delta\alpha$-force increases with decreasing pressure as well as increasing mean free path of gas molecules. For a pressure of mbar or below, the mean free path is approximately at the scale of tens of microns. This means that the photophoretic force is an important force for μm-sized particles at pressures of a few mbar or below. Thus, for a levitated microsphere in an optical trap in low-pressure environments, the motion of sphere can be highly influenced by photophoresis.

3．Motion of heated particle

Typically, due to impurities, the particle in the optical trap will absorb part of the trapping light and convert it into heat. If the gas pressure is low, the interaction between the gas molecules and the particle is insufficient. Then, the energy absorbed by the particle cannot be dissipated and the particle will be in a state of thermal non-equilibrium. In addition, differences in surface roughness and composition will result in variations in the accommodation coefficient over the particle surface. Thus, particle motion is also affected by the randomly oriented photophoretic force $F_{\Delta\alpha}$. In our experiment, we observed a phenomenon that the microspheres can easily escape from the optical trap with no feedback when the pressure in the chamber drops to below a few mbar.

**3.1. Dynamical model in a dual-beam optical trap**

Two most frequently used configurations for optical capturing of micron-sized particles include: upward single beam and counter-propagating beams. The dual-beam optical trap can offer 3D manipulation of particles ranging from hundreds of nanometers to tens of microns in vacuum. It is more suitable for precision sensing in various environments in the future and we employ this configuration in experimental setup. In our study, we choose the case of a dual-beam trap to demonstrate the escaping mechanism of levitated particles in vacuum.

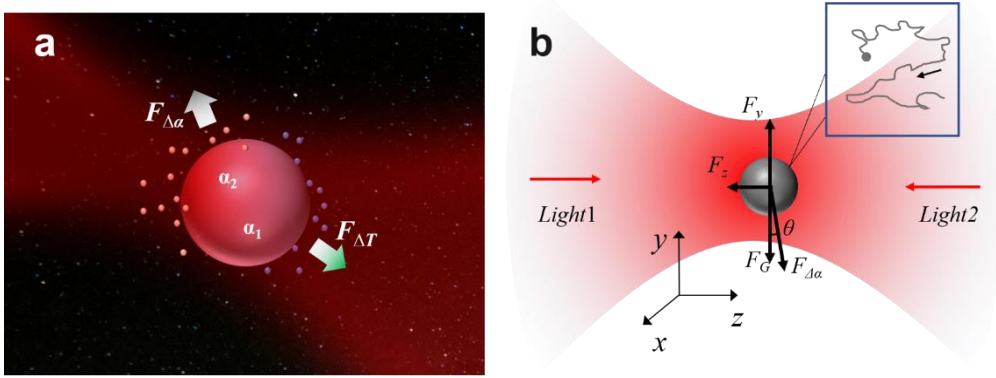

Figure 1. (a) Illustration of the origin of photophoretic forces $F_{\Delta T}$ and $F_{\Delta \alpha}$. A particle having different accommodation coefficients on the two hemispheres ($\alpha_1 > \alpha_2$) experiences a resultant force $F_{\Delta \alpha}$ pointing outward from the surface with the smaller accommodation coefficient $\alpha_2$. Red for "hot" molecules and blue for "cold" molecules. (b) Schematic of a levitated sphere in a dual-beam optical trap. $F_y$, $F_z$ are optical trapping force in y-direction and z-direction respectively and $F_G$ is gravity force. The angle $\theta$ is an indication of the angle between $F_{\Delta \alpha}$ and $F_G$ at a certain time. The red arrows indicate the directions of beam propagation. The inset is a diagram of the example trajectory of the trapped sphere in the radial $xy$-plane.

In order to study the mechanism for loss of the trapped sphere during reducing the ambient pressure without external cooling of sphere's center-of-mass motion, we model the motion of a spherical particle in a dual-beam optical trap. The schematic of the model is shown in Fig.1(b). To simplify the model, we assume that the accommodation coefficient of one half of the sphere surface is $\alpha_1$ and that of the other is $\alpha_2$ (see Fig.1(a)). Since the microspheres used in the experiment are not pure silica spheres, we equivalent it to a relatively simplified model, using α1 and α2 to represent the impure silica spheres. In fact, Δα is hard to be measured on individual particle and there are also differences in $\alpha$ of the same batch of microspheres. The equation of center-of-mass motion can be described classically and is given by Newton's second law [17,19]:

$$\ddot{\boldsymbol{r}}(t) + \Gamma_{CM}\dot{\boldsymbol{r}}(t) = \frac{1}{m}\left[\boldsymbol{F}_{fluct}(t) + \boldsymbol{F}_{det}(\boldsymbol{r},t)\right] \tag{5}$$

where $\boldsymbol{r}$, $m$ represents the position and mass of particle, and $\Gamma_{CM}$ represents the damping rate. $\boldsymbol{F}_{fluct}(t)$ and $\boldsymbol{F}_{det}(\boldsymbol{r}, t)$ represent stochastic forces and deterministic forces, respectively.

In our model, we assume that the temperature of sphere is higher than that of the surroundings. Gas molecules impinge on the sphere surface at temperature $T_{imp}$ and leave at temperature $T_{em}$ ($T_{em} > T_{imp}$). $\Gamma_{imp}$ and $\Gamma_{em}$ are the damping rates for the sphere in connection with the impinging gas and emerging gas respectively[19] and $\Gamma_{CM} = \Gamma_{imp} + \Gamma_{em}$ (see Section 1 of SI). Accordingly, the random fluctuation force

$F_{fluct}(t)$ has two contributions, $F_{imp} = \sqrt{2k_B T_{imp} \Gamma_{imp} m}\xi(t)$ and $F_{em} = \sqrt{2k_B T_{em} \Gamma_{em} m}\xi(t)$, where $k_B$ is Boltzmann constant. Here, $\xi(t)$ encodes a white-noise process, such that $\langle \xi(t) \rangle = 0$, $\langle \xi(t)\xi(t+\tau) \rangle = \delta(\tau)$ [33]. In Fig.1(b), the microsphere is heated by the two counter-propagating laser beams, so the $\Delta T$-forces in axial($z$) direction induced by each beam are in opposite directions and have the same value. The temperature gradient on the sphere remains constant even if the microsphere moves irregularly around the equilibrium point. Thus, these two exactly cancel each other and there is no $\Delta T$-force exerted on the levitated sphere. The direction of gravity is set against $y$ axis and the origin of the coordinate system is at the center of beam waist. The equations of motion for particle in the three directions of the coordinate system read as:

$$m\ddot{z}(t) = F_z + F_{imp}(t) + F_{em-z}(t) + F_{\Delta\alpha-z}(t) - m(\Gamma_{imp} + \Gamma_{em-z})\dot{z}(t) \tag{6}$$

$$m\ddot{x}(t) = F_x + F_{imp}(t) + F_{em-x}(t) + F_{\Delta\alpha-x}(t) - m(\Gamma_{imp} + \Gamma_{em-x})\dot{x}(t) \tag{7}$$

$$m\ddot{y}(t) = F_y - F_G + F_{imp}(t) + F_{em-y}(t) + F_{\Delta\alpha-y}(t) - m(\Gamma_{imp} + \Gamma_{em-y})\dot{y}(t) \tag{8}$$

We predict that the microsphere may escape when the displacement exceeds the linear region, which is confirmed in the following calculations. To simplify the model, we only consider the displacements of the microsphere within the linear region. $F_q$ ($q=x, y, z$) is the corresponding optical force in $q$ direction. Since the $\Delta\alpha$-force has a random direction and is independent of illumination, we assume that the photophoretic force $F_{\Delta\alpha}$ fluctuates on the same time scale as the random forces $F_{imp}(t)$ and $F_{em}(t)$. Then, $F_{\Delta\alpha}(t) = F_{\Delta\alpha}\xi(t)$.

Schematic of our experimental apparatus is shown Fig.2(a). Laser light at a wavelength $\lambda$=1064 nm is split into two horizontal beams and the optical power of each beam is 150 $m$W. The beam waist radius is 5 μm. The focal planes of two beams are overlapping. Fig.2(b) show the experimental and calculated standard deviation (STD) of $x$ displacement for a trapped sphere (a=5 μm) in this optical trap. Here, we assume that the complex refractive index $m=1.45+1\times10^{-4}i$. Then, we can obtain $Q_a$=0.056 by three-dimensional numerical simulation using FDTD methods. The specific description of the simulation is shown in Sections 2 of SI. We substituted the displacement measured in the experiment into the model to calibrate $\Delta\alpha$. By fitting the experimental data to our model, we determined $\Delta\alpha = 3\times10^{-8}$. Since in the experiments, the particles are observed to easily escape at 0.1mbar, we are more concerned about the change of displacement amplitude when the pressure drops from 10 mbar to 0.1mbar, shown in Fig.2(b). The experimental results are in good agreement with the calculated ones in the tendency. The standard deviation of displacement increases as the pressure decreases, and the ratio of that at 0.1 mbar and 10 mbar are about 2.7 and 2.9 for calculated and experimental data respectively.

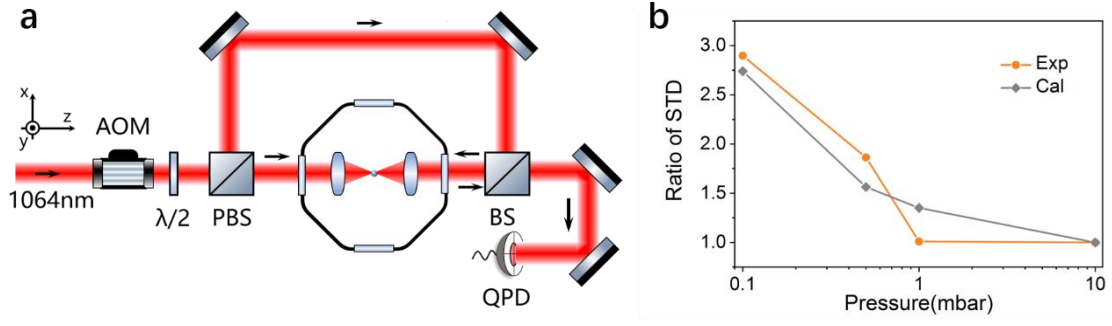

Figure 2. (a)Simplified experimental setup. A sphere (a =5 μm) is levitated in a horizontal dual-beam optical trap. The motion of particle is monitored by quadrant photodetector (QPD). The wavelength of two Gaussian beams is 1064 nm, the waist radius is 5 μm, and the power is 150 mW. (b)Experimental and calculated standard deviation of x-displacement. The ratio of STD is obtained by dividing the standard deviation at a certain pressure by the standard deviation at 10 mbar in each case. The sphere (a =5 μm) is trapped in a dual-beam optical trap. The parameters of simulation refers to experimental parameters.

In Fig.3, we compare the radial motion position trace of a heated particle with or without considering $F_{\Delta\alpha}$. A silica sphere (a=5 μm) with $\Delta\alpha=1.5\times10^{-5}$ is trapped by two equal-power counterpropagating beams at the wavelength of 1064 nm. The power of each beam is 100 mW. The beam waist radius of each beam is 5 μm and the focus points of the two beams are overlapping. The pressure dependence of the particle's x-displacement is demonstrated. For each pressure, the values of STD were averaged over 50 independent simulations that were carried out using the two equations of motion. When the photophoresis mechanism is not considered, term of $F_{\Delta\alpha}$ in equation (7) is missing. The magnitude of the standard deviation of the displacement is on $10^{-8}$ nm. It seems that the sphere is still tightly confined in the optical trap even the sphere is in a thermal non-equilibrium state. However, the sphere tends to move more vigorously at the same pressure if photophoresis is taken into account (red line).

In 2008, Hans Rohatschek [26] gave a simple description of stochastic processes associated with photophoresis. The mean square deviation of the particle random walk reads,

$$\langle x_j^2 \rangle = \frac{2kt}{\gamma}\left(T + \frac{mF_{\Delta\alpha}^2}{3k\gamma^2}\right) \qquad (9)$$

Where $\gamma=m\Gamma_{imp}$ is the damping coefficient in vacuum. Equation (9) indicates that the overall temperature is given by the sum of the actual ambient temperature and a contribution due to the kinetic energy corresponding to the asymptotic velocity $F_{\Delta\alpha}/\gamma$. The $\Delta\alpha$-force just acts as a "amplifier" for the Brownian motion. The effective temperature $T_{eff}=mF_{\Delta\alpha}^2/3k\gamma^2$ is an equivalent describing the photophoretic contribution to stochastic particle displacement as modified Brownian motion

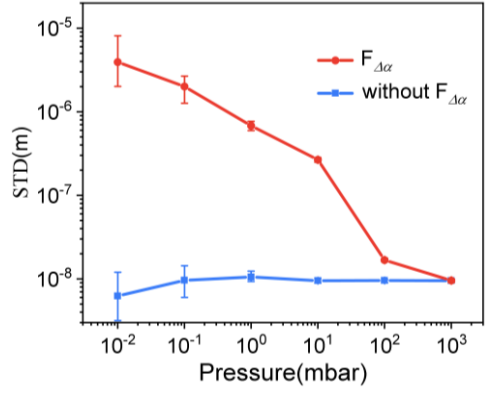

Fig 3. Comparison of particle's *x*-displacement calculated by equation (7) with or without considering $F_{\Delta\alpha}$. Radial motion position(*x*) trace for a trapped sphere(a=5μm) with $\Delta\alpha = 9\times10^{-6}$ in a dual-beam trap. and we assume $T_{em-x}$ = 500K. The pressure dependence of the standard deviation (STD) of the particle's *x*-displacement is shown. For each pressure, the values of STD were averaged over 50 independent simulations. The color bars denote the maximum STD and minimum STD in 50-times calculations.

**3.2. Effective capture region**

Two factors determine whether a particle in the light field can be captured: the range of motion and the region of capture. To further explain the escape process of the particle, we introduce the definition of effective capture region (ECR) proposed by Fu [35]. The effective capture region (ECR) is defined as a criterion for capture of particles, which simply considers particles with zero initial velocity. The initial position $x_0$ with $v_0 = 0$ at time $t_0$ can be considered evolved from a situation where a particle loaded at a previous position $x_p$ with a velocity $v_p \neq 0$ in a previous time, thus the initial position is equivalent to the previous position. If the particle passes through an equivalent position and this equivalent position is also located in the ECR, the particle can be captured. Conversely, particles that do not enter ECR will escape, and particles that enter ECR with high loading velocity will escape afterwards anyway.

In this paper, we focus on the escaping process of a captured particle, so the initial position is at the equilibrium position. Thus, we pay more attention to the extreme value of ECR in the radial direction at z=0. Fig.4 shows simulated the ECR of a dual-beam optical trap (shares the same parameters with Fig.3) for a sphere of a=5μm in 3D at pressures of 10mbar and 0.1mbar (See the simulation of ECR of a vertical single-beam optical trap in Section 3 of SI). Theoretically, the size of the ECR along each axis decreases with pressure because of the decrease in viscous force [35]. Our calculation results support this conclusion as demonstrated in Fig.4. For 10 mbar in Fig.4(a), the viscous force is enough to counteract the kinetic energy from gravitational acceleration, while the optical trapping force in the radial direction plays a minimal role, thus the sphere can be loaded far above the trap radially in a wide range. However, for 0.1 mbar in Fig.4(c), the viscous force becomes much smaller and the optical trapping force makes a major contribution in the loading process, thus decreasing the volume of ECR. The extreme coordinates $x_{extreme}$ at different pressures ranging from 0.1 mbar to $10^3$ mbar are shown in Fig.5(b). The axial location at which the value of $x_{etreme}$ is observed roughly correspond to the axial equilibrium location of the particle. With the pressure decreasing, the extreme value that marks the farthest boundary at which a microsphere can reach while remaining trapped has shrunk. In addition, as the pressure decrease, the difference of ECR range between *x*-direction and *y*-direction become smaller. The ECR range near axial equilibrium location in both directions nearly the same at 0.1mbar. So we take the *x*-direction as an example to analyze the escape mechanism in the following discuss.

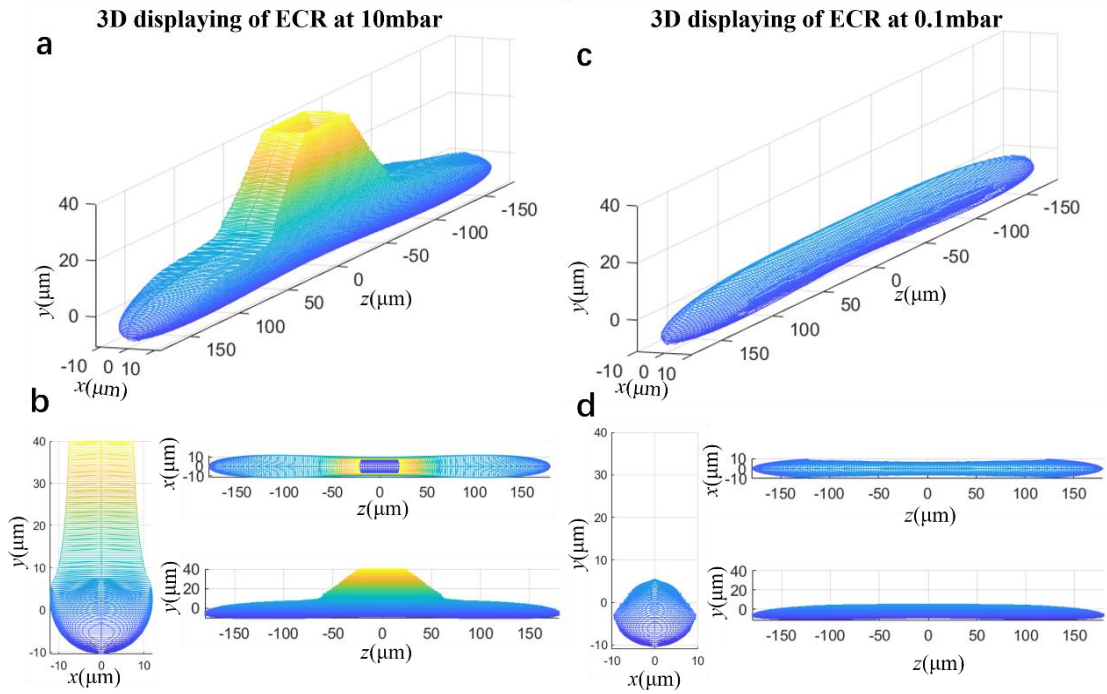

Fig.4. 3D simulation of the ECR of a single-beam trap. 3D display of ECR at (a) 10mbar and (c) 0.1mbar. The ECR in *xy*-plane, *xz*-plane and *yz*-plane at (b) 10mbar and (d) 0.1mbar. Compared with result under the condition of 10mbar, the volume of ECR at 0.1mbar appeared significantly shrinked.

### 3.3. Analysis of escape of heated sphere

Fig.5(a) shows the motion position trace in *x*-direction calculated by Equation (7) for the trapped sphere with $\Delta \alpha = 1.5 \times 10^{-5}$ in a dual-beam trap(See Section 4 of SI for the single-beam optical trap). Other parameters refer to Fig.3 We consider that the sphere is in a state of thermal equilibrium when the pressure is $\geq$ 10 mbar, where the model of two independent heat baths interacting with the trapped particle breaks down. This means the terms $F_{em-x}(t)$ and $M\Gamma_{em-x}\dot{x}(t)$ in Eq. (7) are both equal to zero. As a result, we assume $T_{em} = 0$ under this condition. When the pressure is lower than 10 mbar, the particles will be heated due to the decrease of gas damping. Since the light intensity in our calculation is on the order of $10^9 W/m^2$, we take $T_{em}$=500K here according to Ref[19](See Section 2 of SI for more detail). The axial position of the sphere in the simulations was assumed to be zero axial position. In this paper, we are more concerned with its escape mechanism than with its practical application in sensing or measurement. Therefore, coupling between different axes is not considered in the model. Fig.5(c) shows the trapping force in *x*-direction at zero-axial position. The black line is a polynomial fit for the optical trapping force in the range within orange shadow.

It is obvious that the motion in the radial direction is increasingly more vigorous with the pressure dropping and the contrast is particularly strong for 1 mbar and 0.1 mbar seen in Fig.5(a). The maximum displacement is about 0.65 μm for 1 mbar and 2.65 μm for 0.1 mbar in Fig.5(b). However, there is almost no difference for the motion trajectory at 1 atm, calculated with or without considering photophoretic force shown in Fig.3. This indicates that the photophoretic force plays a major role in the intensification of the movement at pressure of a few mbar. In addition, we can see that the ECR shrinks rapidly with the pressure dropping from 10 mbar to 0.01 mbar in Fig.5(b). At P=0.1 mbar, the extreme value $x_{extreme}$ is about 6.7 μm, which is smaller than the maximum displacement for this pressure. Consequently, the sphere will escape from the optical trap at a certain pressure between 0.1 mbar and 1 mbar. Furthermore, little difference in Δα makes evident change for the motion according to our simulation. Since the impurities [36] and roughness of surface have an important effect on the accommodation coefficient α, higher purity and better sphericity of sphere are expected for better levitation stability.

In this paper, we are more concerned with its escape mechanism than with its practical application in sensing or measurement. Therefore, coupling between different axes is not considered in the model. Fig.5(c) shows the trapping force in $x$-direction at zero-axial position. The black line is a polynomial fit for the optical trapping force in the range within orange shadow.

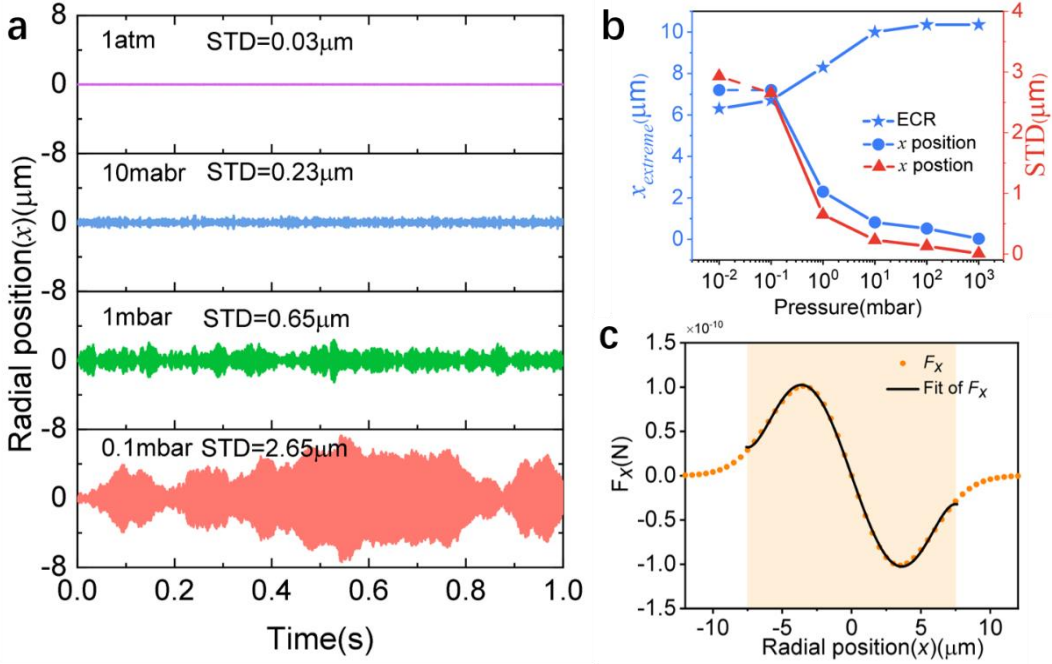

Figure 5. Analysis of sphere escaping from a dual-beam optical trap due to thermal effects. (a) Radial motion position ($x$) trace for a trapped sphere (a =5 μm) with $\Delta\alpha = 2\times10^{-5}$ as a function of pressure: 1 atm, 10 mbar, 1 mbar, 0.1 mbar. The wavelength of beam is 1064 nm, the waist radius is 3 μm, and the power is 100 mW. (b) Left: simulated extreme coordinates $x_{edge}$ of ECR(start) at zero axial position and calculated extreme coordinates of $x$ position(circle) at different pressures ranging from 0.1 mbar to 100 mbar. Right: calculated $x$ displacement standard deviation at different pressures. (c) The radial force of the trapped sphere. The orange shaded area marks the fitting region: $F_x = 7.6\times10^{-11}x^7 - 1.85\times10^{-8}x^5 + 1.5\times10^{-6}x^3 - 4.5\times10^{-5}x$.

In conclusion, the escaping mechanism of particles in an optical trap under vacuum includes two aspects. First, the thermal effects of particles induce a photophoretic force that will enhance the Brownian motions, and this phenomenon becomes increasingly evident during pumping down. Second, the volume of effective capture region shrinks when the pressure decreases. As a result, the range for radial motion to ensure that particles are trapped in the optical trap also shrinks.

### 4．Conclusion

We have derived and tested a dynamic model for the motion of a particle in a thermal non-equilibrium state. The motion of particles was simulated on the basis of dynamic equation in the composite force field. Two types of photophoretic forces: $\Delta T_s$-force and $\Delta\alpha$-force induced by thermal effects have been investigated. It is shown that irregular photophoresis can be described as Brownian motion with increased "effective temperature". We also simulated the ECR of the particle in our model and observed that the ECR tend to shrink apparently at a specific pressure interval. The processes of particle's escape due to thermal effects for this interval was demonstrated and were consistent with the escape phenomenon in the experiment. Our work reveals the escaping mechanism of a heated particle in an optical trap in vacuum and open prospects for increasing the trapping stability of levitated particles in optical trap in vacuum.

**Funding.** Zhejiang Provincial Natural Science Foundation of China under Grant No. LD22F050002; National Natural Science Foundation of China (No.62205290, No. 62075193，No. 62005248); Major Scientific Research Project of Zhejiang Lab; China (No. 2019MB0AD01) and National Program for Special Support of Top-Notch Young Professionals, China (No. W02070390). Center-initiated Research Project of Zhejiang Lab (2022MB0AL02).

**Disclosures.** The authors declare no conflicts of interest.

**Data availability.** Data underlying the results presented in this paper are not publicly available at this time but may be obtained from the authors upon reasonable request.

**Supplemental document.** See Supplement 1 for supporting content.